\documentclass[reqno]{amsart}

\newtheorem{lem}{Lemma}
\newtheorem{thm}{THEOREM}

\theoremstyle{definition}
\newtheorem{definition}{DEFINITION}
\theoremstyle{definition}

\theoremstyle{definition}

\newcommand{\R}{\mathbb{R}}

\newcommand{\be}{\begin{equation}}
\newcommand{\ee}{\end{equation}}
\newcommand{\bea}{\begin{align}}
\newcommand{\eea}{\end{align}}

\newcommand\eps\epsilon

\DeclareMathOperator{\sgn}{sgn}

\begin{document}
\title[BCS Critical Temperature]{The BCS Critical Temperature for Potentials with Negative Scattering Length}

\thanks{\copyright\,2008 by the authors.  This paper may be
  reproduced, in its entirety, for non-commercial purposes. R.S. 
acknowledges partial support by U.S. National Science
Foundation grant PHY-0652356 and by an A.P. Sloan Fellowship}

\author{Christian Hainzl} \address{Christian Hainzl, Departments of
  Mathematics and Physics, UAB, 1300 University Blvd, Birmingham AL
  35294, USA} \email{hainzl@math.uab.edu}

\author{Robert Seiringer} \address{Robert Seiringer, Department of
  Physics, Princeton University, Princeton NJ 08542-0708, USA}
\email{rseiring@princeton.edu}

\date{May 15, 2008}

\begin{abstract}
We prove that the critical temperature for the BCS gap equation is given by 
$$
T_c = \mu \left( \frac 8\pi e^{\gamma -2} + o(1) \right) e^{\pi/(2\sqrt \mu a)}
$$
in the low density limit $\mu\to 0$, with $\gamma$ denoting Euler's constant.
The formula holds for a suitable
class of interaction potentials with negative scattering length $a$ in
the absence of bound states.
\end{abstract}

\maketitle

\section{Introduction and Main Result}

The BCS gap equation \cite{bcs}
\begin{equation}\label{bcse}
\Delta(p) = - \frac 1{(2\pi)^{3/2}} \int_{\R^3} \hat V(p-q) \frac{\Delta(q)}{E(q)} \tanh\frac{E(q)}{2T} dq\,,
\end{equation}
with $E(p)=\sqrt{(p^2-\mu)^2 + |\Delta(p)|^2}$, has played a prominent
role in physics in the fifty years since its introduction. In
(\ref{bcse}), the function $\Delta$ is interpreted as the order
parameter describing paired fermions (Cooper pairs) interacting via a
local pair potential $2 V(x)$, with Fourier transform $\hat V(p) =
(2\pi)^{-3/2} \int_{\R^3} V(x)e^{-ipx} dx $. The (positive) parameters
$T$ and $\mu$ are the temperature and the chemical potential,
respectively. Originally, Eq.~(\ref{bcse}) was introduced to describe
electrons in a crystal, where the interaction is mediated by phonons
and is non-local. Special interest in local potentials of the form
(\ref{bcse}) has recently emerged for the description of ultra-cold
fermionic gases (see, e.g., \cite{zwerger} and references therein). 

It was shown in \cite{HHSS} that the critical temperature for the existence of
non-trivial solutions of the BCS gap equation can be characterized as follows.

\begin{definition}\label{def1}
  Let $\mu>0$ and $V\in L^{3/2}$ be real-valued. Let $K_{T,\mu}$
  denote the multiplication operator in momentum space
$$
K_{T,\mu} = \frac{ |p^2 -\mu|}{\tanh(|p^2-\mu|/(2T))}\,.
$$
The critical temperature for the BCS equation is given by 
\begin{equation}\label{deftc}
\boxed{\ T_c = \inf \left\{ T>0 \, : \, K_{T,\mu} + V \geq 0\right\}\ }
\end{equation}
\end{definition}

It was proved in \cite{HHSS} that Eq.~(\ref{bcse}) has a non-trivial
solution if and only if $T< T_c$. Note that this result gives a linear
criterion for the existence of solutions of the nonlinear equation
(\ref{bcse}).  An analysis of $T_c$ for weak coupling (i.e., $V$
replaced by $\lambda V$ with $\lambda \ll 1$) was given in
\cite{FHNS,HS} (see also \cite{HS2}). In contrast, here we are
interested in the low density limit $\mu \to 0$ at fixed interaction
potential $V$. In this regime, $T_c$ turns out to be related to the
scattering length of $2V$, which can be conveniently defined as
follows.

\begin{definition}\label{def2}
Let  $V\in L^1(\R^3)\cap L^{3/2}(\R^3)$ be real-valued, and let $V(x)^{1/2} = {\rm sgn}(V(x)) |V(x)|^{1/2}$. If $-1$ is not in the spectrum of the Birman-Schwinger operator  $V^{1/2}
\frac 1{p^2} |V|^{1/2}$, then the {\it scattering length} of $2V$ is given by 
\begin{equation}\label{defa}
\boxed{\ a = \frac 1{4\pi} \langle |V|^{1/2}| \frac 1 {1+V^{1/2} \frac 1{p^2} |V|^{1/2}} |V^{1/2}\rangle \ }
\end{equation}
\end{definition}

Note that since $V\in L^{3/2}(\R^3)$ by assumption, the
Birman-Schwinger operator is of Hilbert-Schmidt class by the
Hardy-Littlewood-Sobolev inequality \cite[Thm.~4.3]{LL}. Although it is not self-adjoint, its spectrum is real. The fact that $-1$ is
not in its spectrum means that $p^2 + V$ does not have a zero
eigenvalue or resonance. Eq.~(\ref{defa}) is the natural definition of
the scattering length for integrable potentials.  In the appendix, we
shall explain why this definition coincides with the usual concept
found in quantum mechanics textbooks. For an alternative definition
allowing for more singular local behavior but assuming compact support
of $V$ and the absence of bound states see \cite[Appendix~A]{LY}.

With the aid of Definitions~\ref{def1} and~\ref{def2} we can now state our main theorem.

\begin{thm}\label{main}
  Assume that $V(x)(1+|x|)\in L^1(\R^3)\cap L^{3/2}(\R^3)$ is
  real-valued. Assume that the spectrum of $V^{1/2} \frac 1{p^2}
  |V|^{1/2}$ is contained in $(-1,\infty)$, and that the scattering
  length $a$ in (\ref{defa}) is negative.  Then the critical
  temperature (\ref{deftc}) satisfies
\begin{equation}\label{eq:main}
\boxed{\ \lim_{\mu \to 0}  \left(\ln\frac \mu {T_c} + \frac \pi {2\sqrt\mu\, a}\right) =  2-\gamma -\ln\frac 8\pi \ }
\end{equation}
with $\gamma\approx 0.577$ denoting Euler's constant. 
\end{thm}

In other words,
$$
T_c = \mu \left( \frac 8\pi e^{\gamma -2} + o(1) \right) e^{\pi/(2\sqrt \mu a)}
$$
as $\mu\to 0$. This formula is well-known in the physics literature
\cite{gorkov,leg,NS,PB}. The operator $V^{1/2} \frac 1{p^2} |V|^{1/2}$
having spectrum in $(-1,\infty)$ implies, in particular, that $p^2 +V$
does not have any bound states. The proof of Theorem~\ref{main} shows
that the condition that $V(x)|x|$ is integrable at infinity is indeed
optimal; i.e., formula (\ref{eq:main}) will in general not hold if $V$
decays slower than $|x|^{-4}$ at infinity.

In the next section we shall prove Theorem~\ref{main}. Our analysis of
the operator $K_{T,\mu}+V$ in (\ref{deftc}) is similar in spirit to
the spectral analysis of Schr\"odinger operators in two dimensions in
\cite{simon,simon2}.

\section{Proof}
Let $-1/\lambda$ denote the smallest eigenvalue of  $V^{1/2}\frac 1{p^2}|V|^{1/2}$. By
assumption, $\lambda>1$. It follows that for this value of $\lambda$,
$p^2 + \lambda V \geq 0$.

Since $\tanh t \leq \min\{1,t\}$ for $t\geq 0$ it is easy to see that
$K_{T,\mu} \geq \lambda^{-1} ( |p^2-\mu| + 2T(\lambda-1) )$ for any
$\lambda\geq 1$. With the choice of $\lambda$ as above, we conclude
that $K_{T,\mu} + V \geq \lambda^{-1} ( p^2 + \lambda V - \mu +
2T(\lambda-1)) \geq \lambda^{-1}( - \mu + 2T(\lambda-1))$. From this
bound and (\ref{deftc}) it follows immediately that $T_c\leq
\mu/(2(\lambda-1))$. Hence we restrict our attention to the case
$T\leq D\mu$ in the following, with $D=1/(2(\lambda-1))>0$. Only the
existence of such a $D$, and not its value, will be important in the
following.

According to the Birman-Schwinger principle (see
\cite[Lemma~1]{FHNS}), $T_c$ is determined by the fact that for
$T=T_c$ the smallest eigenvalue of
$$
B_T = V^{1/2}\frac 1 {K_{T,\mu}} |V|^{1/2}
$$ 
equals $-1$. If the spectrum of $B_T$ is contained in $(-1,\infty)$
for any $T>0$, then $T_c=0$.  Alternatively, $T_c$ is the largest $T$
such that $1+B_T$ has an eigenvalue 0.

We decompose the Birman-Schwinger operator $B_T$ as 
$$
B_T= V^{1/2} \frac 1{K_{T,\mu}} |V|^{1/2} = V^{1/2} \frac 1{p^2}
|V|^{1/2} + m_\mu(T) |V^{1/2}\rangle\langle |V|^{1/2}| + A_{T,\mu} \,,
$$
with
$$
m_\mu(T) = \frac 1{(2\pi)^3} \int_{\R^3} \left( \frac 1{K_{T,\mu}(p)} - \frac 1 {p^2} \right) dp\,.
$$
Explicitly, $A_{T,\mu}$ is the operator with integral kernel
$$
A_{T,\mu}(x,y) = V(x)^{1/2} |V(y)|^{1/2}  \frac1{(2\pi)^3}\int_{\R^3} \left( e^{ip(x-y)} - 1\right) \left( \frac 1{K_{T,\mu}(p)} - \frac 1 {p^2} \right) dp\,.
$$
We note that for small $\mu$
\begin{equation}\label{defm}
m_\mu(T) = \frac {\sqrt{\mu}}{2\pi^2} \left( \ln\frac \mu T + \gamma -2 +\ln \frac 8 \pi + o(1) \right)
\end{equation}
uniformly in $T$ for $T\leq D\mu$. This was shown in \cite[Lemma~1]{HS}.

\begin{lem}\label{hsb}
Under the assumption that $\int_{\R^3} |V(x)|(1+ |x|) dx < \infty$,
$$
\lim_{\mu\to 0}  \sup_{0<T\leq D\mu}\frac 1{ \mu^{1/4} m_\mu(T)}  \|A_{T,\mu}\|_2 = 0\,. 
$$
\end{lem}

Here, $\|\, \cdot \,\|_2$ denotes the Hilbert-Schmidt norm. 

\begin{proof}
Performing the angular integration, we can write the kernel of $A_{T,\mu}$ as 
\begin{equation}\label{amt}
A_{T,\mu}(x,y) = V(x)^{1/2} |V(y)|^{1/2} \frac 1{2\pi^2} \int_{0}^\infty \left( \frac{\sin p|x-y|}{p|x-y|} - 1\right) \left( \frac 1{K_{T,\mu}(p)} - \frac 1 {p^2} \right) p^2 dp\,.
\end{equation}
We note that $|b^{-1} \sin b -1| \leq C \min\{b^2,1\} \leq C b^\alpha$ for all $0\leq \alpha\leq 2$ and $b>0$. Hence we can bound, for any $Z>0$, 
\begin{align}\nonumber
\left| \frac{\sin p|x-y|}{p |x-y|} - 1\right| &\leq C \left[ p^2 Z^2 \theta(Z-|x-y|) + \sqrt{p|x-y|} \theta(|x-y|-Z)\right] \theta(2\mu - p^2) \\ & \quad  \nonumber+ C \sqrt{p|x-y|} \theta(p^2-2\mu) \,.
\end{align}
The kernel of $A_{T,\mu}$ is thus bounded by 
\begin{align}\nonumber
& |A_{T,\mu}(x,y)| \\ \nonumber
&\leq |V(x)|^{1/2} |V(y)|^{1/2}  Z^2  \frac C{2\pi^2} \int_{0}^{\sqrt{2\mu}} \left|  \frac 1{K_{T,\mu}(p)} - \frac 1 {p^2} \right| p^4 dp \\ \nonumber  & \quad + |V(x)|^{1/2} |V(y)|^{1/2} |x-y|^{1/2} \theta(|x-y|-Z) \frac C{2\pi^2} \int_{0}^{\sqrt{2\mu}} \left|  \frac 1{K_{T,\mu}(p)} - \frac 1 {p^{2}} \right| p^{5/2} dp \\ \nonumber & \quad + |V(x)|^{1/2} |V(y)|^{1/2} |x-y|^{1/2} \frac C{2\pi^2} \int_{\sqrt{2\mu}}^{\infty} \left|  \frac 1{K_{T,\mu}(p)} - \frac 1 {p^2} \right| p^{5/2} dp
\end{align}
It is not difficult to see that the $p$ integral on the first line is bounded by (a constant times) $ \mu m_\mu(T)$ for $T\leq D\mu$. Similarly, the integral on the second line is bounded by $\mu^{1/4} m_\mu(T)$. Finally, the last integral is bounded by $\mu^{3/4}$. Since by assumption $\int_{\R^3} |V(x)| dx < \infty$ and $\int_{\R^6} |V(x)||V(y)||x-y| dx dy<\infty$, we conclude that 
$$
\limsup_{\mu \to 0}   \frac 1{\mu^{1/4} m_\mu(T)}  \|A_{T,\mu}\|_2 \leq C \left( \int_{|x-y|\geq Z}  |V(x)| |V(y)| |x-y| dx dy  \right)^{1/2}
$$
for some constant $C>0$, uniformly in $T$ for $T\leq D\mu$. 
Since $Z$ was arbitrary, this proves the claim.
\end{proof}

Since $1+V^{1/2}p^{-2}|V|^{1/2}$ is invertible by assumption, we can write
$$
1+B_T = \left( 1+V^{1/2} \frac 1{p^2} |V|^{1/2}\right) \left( 1 + \frac{m_\mu(T)}{1+V^{1/2}p^{-2}|V|^{1/2}}\left( |V^{1/2}\rangle\langle |V|^{1/2}| + \frac{A_{T,\mu}}{m_\mu(T)}\right) \right)\,.
$$
Hence $T_c$ is the largest $T$ such that 
$$
 \frac{m_\mu(T)}{1+V^{1/2}p^{-2}|V|^{1/2}}\left( |V^{1/2}\rangle\langle |V|^{1/2}| + \frac{A_{T,\mu}}{m_\mu(T)}\right)
$$
has an eigenvalue $-1$. According to Lemma~\ref{hsb}, the operator $A_{T,\mu}/m_\mu(T)$ is small for small $\mu$, uniformly in $T$. Moreover, $m_\mu(T)$ is monotone decreasing in $T$ and the rank one operator
$$
\frac 1 {1+V^{1/2}p^{-2}|V|^{1/2}}  |V^{1/2}\rangle\langle |V|^{1/2}|
$$ 
has an eigenvalue $4\pi a$ according to Definition~\ref{def2}, which is
negative by assumption. Hence simple perturbation implies that
$$
\lim_{\mu\to 0} m_\mu(T_c) = - \frac 1{4\pi a}\,.
$$
In particular, $T_c>0$ for small enough $\mu$.

We conclude that $m_\mu(T_c)$ is order one as $\mu\to 0$. With the aid of 
Lemma~\ref{hsb} we thus see that for $T=T_c$, $1+V^{1/2}p^{-2}|V|^{1/2}+
A_{T,\mu}$ is invertible for small enough $\mu$. This implies that
$T_c$ is determined by the fact that the smallest eigenvalue of
\begin{align}\nonumber
&1+ B_T= \\ \nonumber & \left( 1 + V^{1/2} \frac 1{p^2} |V|^{1/2} +A_{T,\mu} \right) \left( 1 + \frac { m_\mu(T)}{  1 + V^{1/2} \frac 1{p^2} |V|^{1/2} + A_{T,\mu}}  |V^{1/2}\rangle\langle |V|^{1/2}|\right)
\end{align}
is zero. Since the first factor does not have a zero eigenvalue, the second does for $T=T_c$. In other words,
\begin{equation}\label{toinsert}
-\frac 1{ m_\mu(T_c)} = \langle |V|^{1/2}| \frac 1{  1 + V^{1/2} \frac 1{p^2} |V|^{1/2} + A_{T_c,\mu} } |V^{1/2}\rangle    \,.
\end{equation}

In order to evaluate this expression, we expand 
\begin{align}\nonumber
& \frac 1{  1 + V^{1/2} \frac 1{p^2} |V|^{1/2} + A_{T,\mu} } \\ \nonumber
& =  \frac 1{  1 + V^{1/2} \frac 1{p^2} |V|^{1/2} }  - \frac 1{  1 + V^{1/2} \frac 1{p^2} |V|^{1/2} }A_{T,\mu}  \frac 1{  1 + V^{1/2} \frac 1{p^2} |V|^{1/2} }  \\ \nonumber &\quad +  \frac 1{  1 + V^{1/2} \frac 1{p^2} |V|^{1/2} }A_{T,\mu}  \frac 1{  1 + V^{1/2} \frac 1{p^2} |V|^{1/2} + A_{T,\mu} } A_{T,\mu}  \frac 1{  1 + V^{1/2} \frac 1{p^2} |V|^{1/2}  }\,.
\end{align}
The first term, when inserted in (\ref{toinsert}), yields $1/(4\pi a)$. The last term term goes to zero faster than $\mu^{1/2}$ since, according to Lemma~\ref{hsb}, $\|A_{T_c,\mu}\|\ll \mu^{1/4}$. Inserting the second term into (\ref{toinsert}) yields $\langle f| (\sgn V)A_{T_c,\mu}|f\rangle$ with
\begin{equation}\label{deff}
  |f\rangle =  \frac 1{ 1 + V^{1/2} \frac 1{p^2} |V|^{1/2}} | V^{1/2}\rangle\,.
\end{equation}

\begin{lem}\label{lem2}
Assume that $f(x) |V(x)|^{1/2}(1+|x|) \in L^1(\R^3)$. Then
\begin{equation}\label{eq:lem2}
\lim_{\mu\to 0} \sup_{0<T\leq D\mu} \frac{1}{\mu^{1/2}m_\mu(T)}\, |\langle f| (\sgn V)A_{T,\mu}|f\rangle|=0\,.
\end{equation}
\end{lem}

\begin{proof}
Using (\ref{amt}) we can write $\langle f| (\sgn V)A_{T,\mu}|f\rangle$ as
$$
\int_{\R^6} \overline{f(x)} |V(x)|^{1/2} f(y) |V(y)|^{1/2}\!\!\int_{0}^\infty \left( \frac{\sin p|x-y|}{p|x-y|} - 1\right) \left( \frac 1{K_{T,\mu}(p)} - \frac 1 {p^2} \right) \frac{p^2 dp}{2\pi^2} dx dy\,.
$$
Proceeding as in the proof of Lemma~\ref{hsb}, but with $\alpha=1$ in place of $\alpha=1/2$ in the second term, we conclude that 
\begin{align}\nonumber
& \left|\int_0^\infty  \left( \frac{\sin p|x-y|}{p|x-y|} - 1\right) \left( \frac 1{K_{T,\mu}(p)} - \frac 1 {p^2} \right) p^2 dp\right| \\ \nonumber
&\leq  Z^2  C \int_{0}^{\sqrt{2\mu}} \left|  \frac 1{K_{T,\mu}(p)} - \frac 1 {p^2} \right| p^4 dp \\ \nonumber  & \quad + |x-y| \theta(|x-y|-Z)  C \int_{0}^{\sqrt{2\mu}} \left|  \frac 1{K_{T,\mu}(p)} - \frac 1 {p^{2}} \right| p^{3} dp \\ \nonumber & \quad + |x-y|^{1/2} C  \int_{\sqrt{2\mu}}^{\infty} \left|  \frac 1{K_{T,\mu}(p)} - \frac 1 {p^2} \right| p^{5/2} dp\,.
\end{align}
The first and second integrals are bounded by $\mu m_\mu(T)$ and
$\sqrt\mu m_\mu(T)$, respectively. The last integral is bounded by
$\mu^{3/4}$. Since $Z$ is arbitrary, the same argument as in the proof of Lemma~\ref{hsb} leads to (\ref{eq:lem2}). 
\end{proof}

If $f$ defined in (\ref{deff}) satisfies the assumption of Lemma~\ref{lem2}, we conclude that $\langle f| (\sgn V) A_{T_c,\mu}|f\rangle \ll \mu^{1/2}$ as $\mu \to 0$. In particular, this implies that   
$$
\lim_{\mu \to 0} \frac 1{\sqrt\mu} \left( m_\mu(T_c) + \frac 1{4\pi a} \right) = 0\,.
$$
Using (\ref{defm}) this proves (\ref{eq:main}).

Thus it remains to show that $f(x) |V(x)|^{1/2}(1+|x|) \in L^1(\R^3)$, with $f$ defined in (\ref{deff}). For this purpose, let $g\in L^\infty(\R^3)$, and consider the expression
\begin{equation}\label{expr}
\int_{\R^3} \overline{g(x)} |V(x)|^{1/2} (1+|x|) f(x) dx = \langle g |V|^{1/2}| (1+|x|)  \frac 1{ 1 + V^{1/2} \frac 1{p^2} |V|^{1/2}} | V^{1/2}\rangle \,.
\end{equation}
We write 
$$
 \frac 1{ 1 + V^{1/2} \frac 1{p^2} |V|^{1/2}}  = 1 - V^{1/2} \frac 1{p^2} |V|^{1/2}  \frac 1{ 1 + V^{1/2} \frac 1{p^2} |V|^{1/2}} \,.
$$
For the contribution of $1$ to (\ref{expr}) we use 
\begin{equation}\label{cont1}
\left| \langle g |V|^{1/2}| (1+|x|)  | V^{1/2}\rangle\right| \leq \|g\|_\infty \int_{\R^3} |V(x)| (1+|x|) dx \,.
\end{equation}
The remaining term is then bounded by
\begin{align}\nonumber 
& \| g|V|^{1/2}(1+|x|)^{1/2}\|_2 \, \| V^{1/2}\|_2 \\ \nonumber & \quad \times \left\| (1+ V^{1/2} p^{-2} |V|^{1/2} )^{-1} \right\|\,  \left\| (1+|x|)^{1/2} V^{1/2}\frac 1{p^2} |V|^{1/2} \right\|  \,.
\end{align}
The first norm is bounded by $\|g\|_\infty (\int |V(x)|(1+|x|) dx)^{1/2}$. The second is just $\|V\|_1^{1/2}$. The first operator norm in the second line is finite by assumption. Finally, the last operator norm can be bounded by the Hilbert-Schmidt norm, whose square is 
$$
\int_{\R^6} (1+|x|) |V(x)| \frac{1}{|x-y|^2} |V(y)| dx dy \leq C \|V\|_{3/2} \|(1+|x|)V\|_{3/2}\,.
$$
The last inequality follows from the Hardy-Littlewood-Sobolev
inequality \cite[Thm.~4.3]{LL}; the right side is finite by our assumptions on
$V$. This shows that (\ref{expr}) is bounded by (a constant times) $\|g\|_\infty$, and
hence $f(x)|V(x)|^{1/2}(1+|x|)\in L^1$. This completes the proof of Theorem~\ref{main}.

\appendix
\section{Scattering Length}

In the following, we shall explain why the definition (\ref{defa}) of the scattering length coincides with the usual textbook definition. There, one considers the solution $\psi$ of the zero-energy scattering equation
$$
-\Delta \psi + V \psi = 0
$$
subject to the boundary condition $\lim_{|x|\to\infty} \psi(x) =
1$. The scattering length $a$ is then determined by the asymptotic
behavior $\psi(x) \approx 1 - a/|x|$ for large $|x|$. For a rigorous
definition for potentials with compact support and in the absence of
bound states see \cite[Appendix~A]{LY}.

Assuming that we have such a solution $\psi$ with $\psi-1 \approx a/|x|$ for large $|x|$, we have  
$$
\frac 1{p^2} \Delta\psi = \frac 1{p^2} \Delta(\psi-1) = 1-\psi  \,.
$$
This can be rewritten as 
$$
V^{1/2} \frac 1 {p^2} V \psi = V^{1/2} \left(1- \psi\right), 
$$
or 
$$
\left( V^{1/2} \frac 1{p^2} |V|^{1/2}  - \frac {|V^{1/2}\rangle\langle |V|^{1/2}|}{\int V \psi}+1\right)  V^{1/2} \psi = 0 \,.
$$
Moreover, $\int V\psi = \int\Delta \psi = 4\pi a$, which follows from integration by parts, using that $|\nabla \psi|\approx a/|x|^2$ for large $|x|$.

Under the assumption that  $1+ V^{1/2} p^{-2} |V|^{1/2}$ is
invertible, we thus have
$$
\left( 1 + V^{1/2} \frac 1{p^2} |V|^{1/2}\right) \left( 1 - \frac 1 {1+V^{1/2} \frac 1{p^2} |V|^{1/2}} \frac {|V^{1/2}\rangle\langle |V|^{1/2}|}{4\pi a}\right) V^{1/2} \psi = 0\,.
$$
Hence the rank one operator 
$$
\frac 1 {1+V^{1/2} \frac 1{p^2} |V|^{1/2}} \frac {|V^{1/2}\rangle\langle |V|^{1/2}|}{4\pi a}
$$
 has an eigenvalue 1, which implies that 
$$
\langle |V|^{1/2}| \frac 1 {1+V^{1/2} \frac 1{p^2} |V|^{1/2}} |V^{1/2}\rangle = 4\pi a\,.
$$

\end{document}